\documentclass[11pt,a4paper]{article}   

\usepackage{amsmath,amsfonts,amssymb}   
\usepackage{graphicx}   

\usepackage[latin1]{inputenc}     
\usepackage[T1]{fontenc}
\usepackage{authblk} 

\usepackage{float}   
\usepackage{soul, color} 
\usepackage{xcolor}

\usepackage{times}

\usepackage[a4paper]{geometry} 
\usepackage{ragged2e}
\justifying

\graphicspath{{./images/}}

\definecolor{lightblue}{rgb}{0.8,0.9,1} 

\begin{document}

\title{Heads or tails in zero gravity:\\ an example of a classical contextual "measurement"}
\author[1]{Alexandre Gondran\thanks{alexandre.gondran@enac.fr}}
\author[2]{Michel Gondran\thanks{michel.gondran@polytechnique.org}}
\affil[1]{{\small \'Ecole Nationale de l'Aviation Civile, Toulouse, France}}
\affil[2]{{\small Académie Européenne Interdisciplinaire des Sciences, Paris, France}}

\date{}

\maketitle

\begin{abstract}

Playing the game of heads or tails in zero gravity demonstrates that there exists a contextual "measurement" in classical mechanics. When the coin is flipped, its orientation is a continuous variable. However, the "measurement" that occurs when the coin is caught by clapping two hands together gives a discrete value (heads or tails) that depends on the context (orientation of the hands). It is then shown that there is a strong analogy with the spin measurement of the Stern-Gerlach experiment, and in particular with Stern and Gerlach's sequential measurements.  Finally, we clarify the analogy by recalling how the de Broglie-Bohm interpretation simply explains the spin "measurement".

\end{abstract}



\section{Introduction}

Understanding measurement in quantum mechanics has always been a major problem. This is especially the case of the "measurement" of the spin in the Stern-Gerlach experiment, which is considered the archetype of a so-called contextual measurement. In Modern Quantum Mechanics \cite{Sakurai2011}, Jun John Sakurai calls attention to this point in his opening remarks: "\textit{This experiment illustrates in a dramatic manner the necessity for a radical departure from the concepts of classical mechanics. [...] A solid understanding of problems involving two-state systems will turn out be rewarding to any serious student of quantum mechanics.}"
Many physicists (primarily those related to the Copenhagen school) have interpreted this experiment as if "\textit{the measurement creates reality}" or more explicitly, as Albert Einstein famously put it, as if  "\textit{the moon doesn't exist when no one looks at it}".
The purpose of this article is to show that contextual measurements also exist in classical mechanics and that they are not mysterious. 
Section~\ref{sect:zero_gravity} defines the specific case of heads or tails in zero gravity, which presents a certain analogy with the measurement of spin in the Stern-Gerlach experiment. This "measurement" corresponds in fact to the orientation of a continuous variable to a discrete two-state value. 
This analogy continues with Stern and Gerlach's sequential measurements in section~\ref{sect:Sequential}. 
Finally, section~\ref{sect:dBB} clarifies the analogy by recalling how the de Broglie-Bohm interpretation \cite{deBroglie1927, Bohm1952} simply explains the spin "measurement"
and we conclude in section~\ref{sect:conclusion}.

\section{Heads or tails in zero gravity}
\label{sect:zero_gravity}
Have you ever tossed a coin in space? You'll run into a small problem once the coin has been tossed into the air: it won't drop into your palm but will remain twirling in the air. The absence of gravity adds a complication to coin tossing. New rules are required when playing heads or tails in space: the tosser has to clap his hands together so that the coin is caught between his right and left palms. Then he will ask his opponent to choose heads or tails. The coin tosser then withdraws his hands. The coin still does not fall, it remains motionless in weightlessness. The side of the coin that was on the tosser-clapper's right palm designates the winner. It is understood in this experiment that the heads/tails orientation of the coin is not determined before the measurement (the clapping together of hands). The orientation of the coin is a variable with non-predetermined values (before measurement), unlike the mass, color, value or size of the coin which are variables with predetermined values. This is what is described as a contextual measurement, i.e. the measurement of a variable with a continuous value (orientation of the coin) that only gives discrete values (heads or tails) depending on the context (the orientation of the hands).

\paragraph*{Analogy with Stern-Gerlach spin measurement}

\begin{figure}[h!]
\begin{center}
 \includegraphics[width=0.6\textwidth]{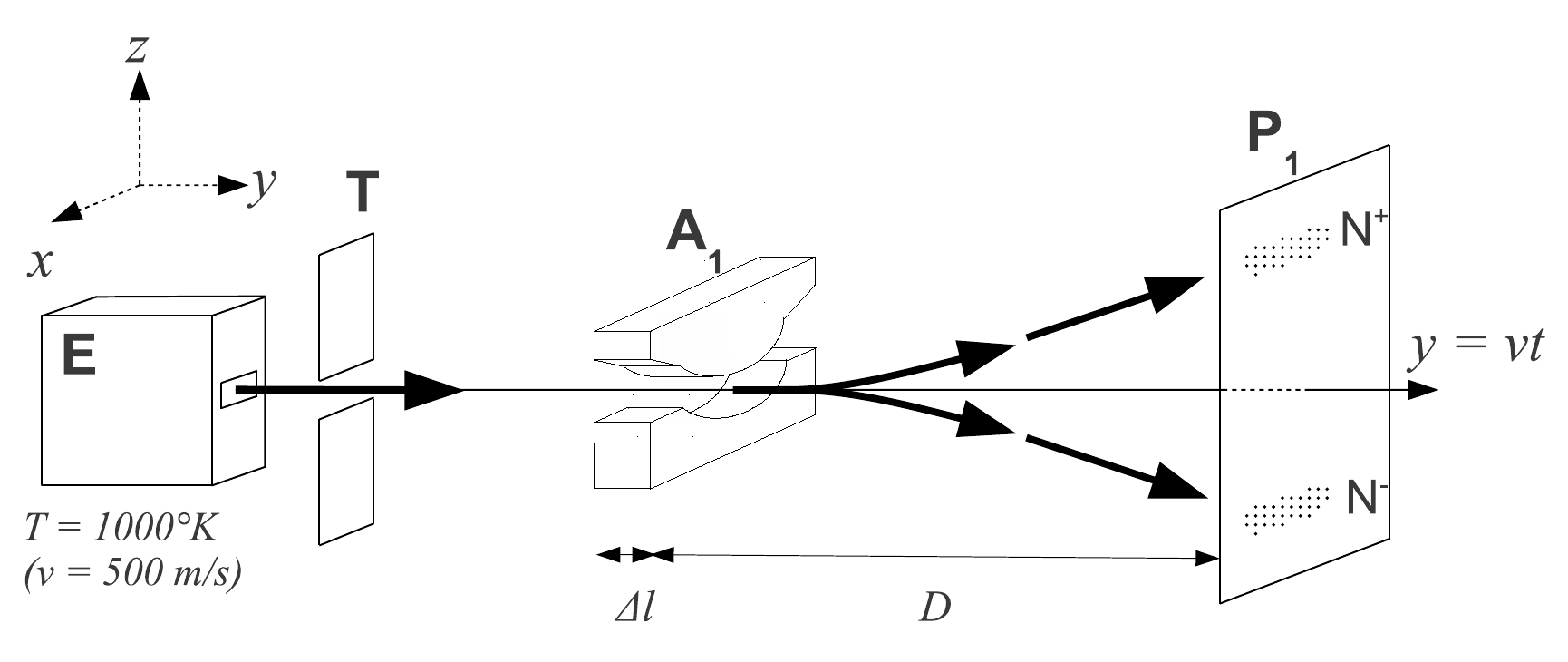}
\caption{\label{fig:SG_schema} Schematic configuration of the historic Stern-Gerlach experiment: a beam of silver atoms~\cite{Gerlach1922}, from the source (oven \textbf{E} , collimator hole \textbf{T}) passes through an inhomogeneous magnetic field (magnet $\mathbf{A_1}$), then separates into two distinct beams before condensing on the screen $\mathbf{P_1}$ on two distinct spots of identical intensity $N^+$ et $N^-$.}
\end{center}
\end{figure}

In quantum mechanics, the orientation of the spin of a particle is also a variable whose values are non-predetermined before measurement. When measuring the spin; let us consider the example of a 1/2 spin particle passing through a Stern-Gerlach type measuring device, that is to say a very powerful magnetic field (cf.~figure~\ref{fig:SG_schema}), there can only be two possible results: either spin up (i.e. a deviation of the particle upwards with respect to the Stern-Gerlach device) or spin down (i.e. a deviation of the particle downwards).

This is the analog of the measurement of heads or tails by clapping hands in space. In both cases, it is possible to say that the orientation of the coin (or the spin) before the measurement is in a superposition of the state of heads and the state of tails (respectively state of up and state of down); mathematically a superposition of states is simply a state that is linear combination of several states. These are variables with continuous values (all superpositions are possible) but whose measurement produces only discrete values.

\section{Sequential measurements of heads or tails}
\label{sect:Sequential}
It is interesting to do measurements in space because it is possible to extend the spin up/down analogy of a particle with the heads/tails orientation of a coin. Indeed, to measure spin, it is possible to orient the measuring instrument (the air gap of the Stern-Gerlach device) in any direction in space; the direction $\widehat{z}$ classically denotes the vertical direction and $\widehat{x}$ and $\widehat{y}$ the other two horizontal directions.
\begin{figure}[h!]
\begin{center}
 \includegraphics[width=0.45\textwidth]{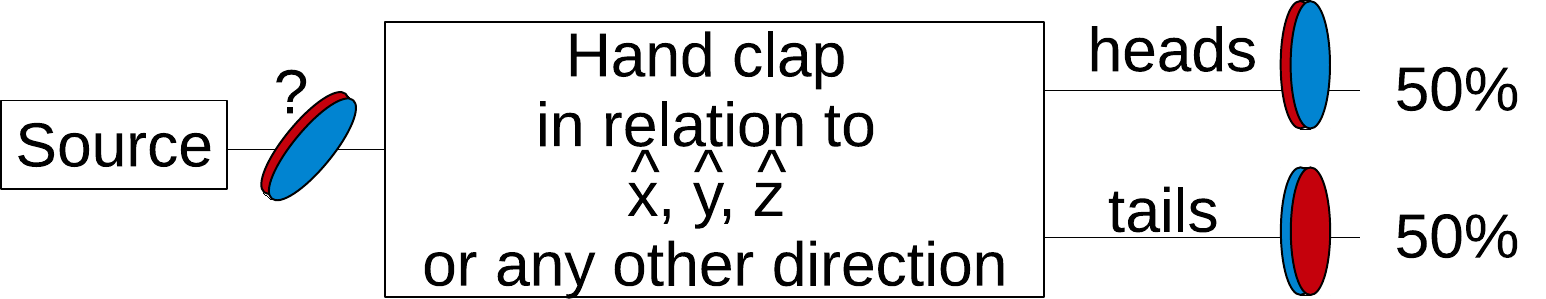}
\caption{\label{fig:1appareil} Measuring the orientation of a coin by clapping hands.}
\end{center}
\end{figure}

In the coin toss game on Earth, the heads or tails orientation is only measured on the vertical axis because of gravity. In zero gravity, there is no longer any specific direction and it is possible to orient the measuring device (i.e. the clapping hands) in all possible directions, and therefore also the measurement of the spin. This could be described as a game of heads or tails in three dimensions. The experiment is schematically represented in Figure~\ref{fig:1appareil}, where it is shown that there is a 50/100  chance of obtaining the heads orientation and a 50/100  chance of obtaining the tails orientation, regardless of the orientation of the measuring device. There is therefore a 50/50 superposition of the states of heads and tails.

\begin{figure}[h!]
\begin{center}
 \includegraphics[width=0.6\textwidth]{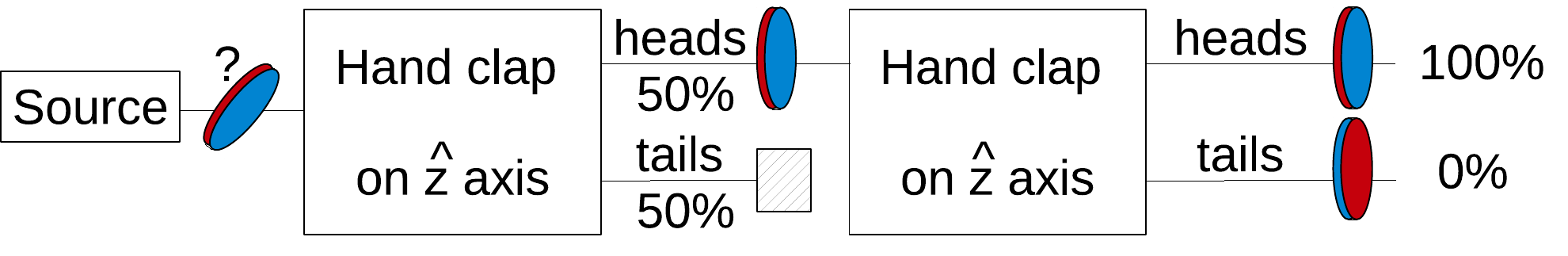}
\caption{\label{fig:2appareils_a} Two successive measurements of the orientation of the coin with measuring devices oriented in the same direction.}
\end{center}
\end{figure}

If, after measuring the coin in the heads state on the $\hat{z}$ axis, a second measurement is made in the same direction, as shown in Figure~\ref{fig:2appareils_a}~, and the hands are again clapped on the $\hat{z}$ axis, it is obvious that the measurement of the orientation once again results in heads.

\begin{figure}[h!]
\begin{center}
 \includegraphics[width=0.6\textwidth]{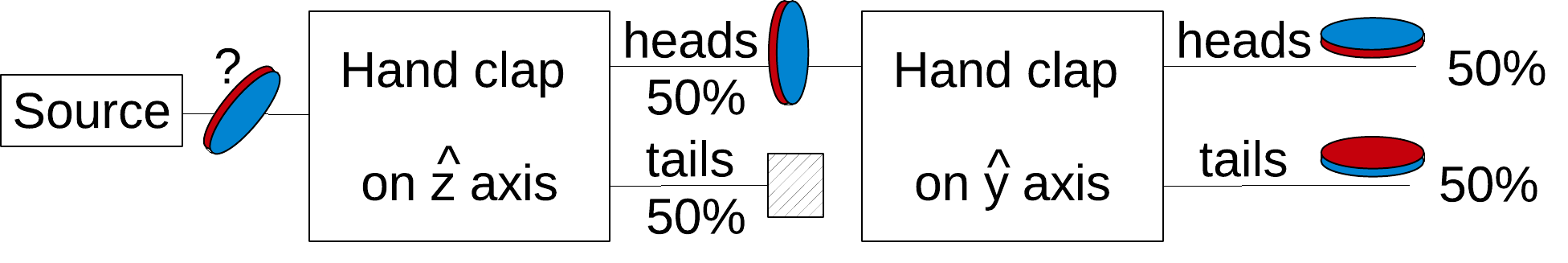}
\caption{\label{fig:2appareils_b}Two successive measurements of the orientation of the coin with measuring devices oriented in perpendicular directions.}
\end{center}
\end{figure}

A second measurement is now made, but on the $\hat{y}$ axis. The first measurement was a vertical clapping of the hands: the coin was forced to orient itself vertically in relation to heads and tails (it is assumed that heads was found, for example); after this measurement, it no longer spins but remains oriented vertically. Now a second measurement is made by clapping hands horizontally this time (see Figure~\ref{fig:2appareils_b}). The coin is once again forced to orient itself but horizontally this time.  Since it is on its side, due to the horizontal orientation, it has a 50 \% chance of being measured heads and a 50\% chance of being measured tails. One final experiment is conducted to understand the behavior of a variable with non-predetermined values.

\begin{figure}[h!]
\begin{center}
 \includegraphics[width=0.8\textwidth]{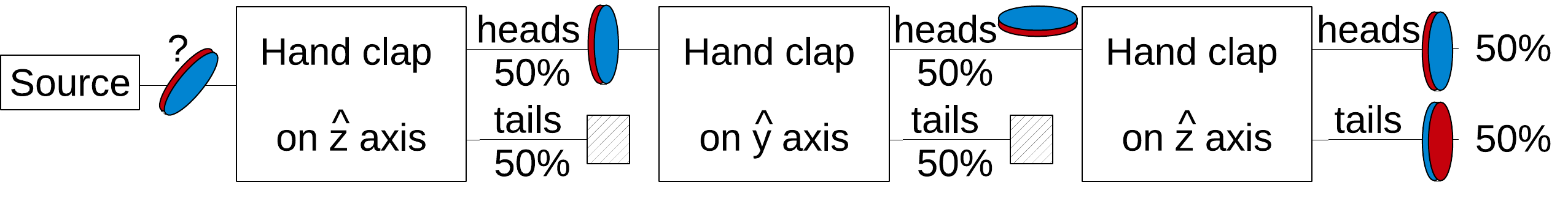}
\caption{\label{fig:3appareils} Three successive measurements of the orientation of the coin with measuring devices oriented respectively in relation to $\hat{z}$ then $\hat{y}$ then again $\hat{z}$.}
\end{center}
\end{figure}

 As shown in Figure~\ref{fig:3appareils}, after the first two measurements in relation to $\hat{z}$ and $\hat{y}$, a third measurement is carried out, again in relation to $\hat{z}$. After the second measurement, the coin floats horizontally in the air and the third hand clap straightens it again. The first measurement in this same direction resulted in heads, but now there is a 50 \% chance that its will be heads again and 50 \% it will be tails.

 The latter experiment shows that a variable with non-predetermined values of a given object can be measured in the same direction and not result in the same value. The four experiments (Figures~\ref{fig:1appareil},~\ref{fig:2appareils_a},~\ref{fig:2appareils_b} and \ref{fig:3appareils}) are characteristic of variables with non-predetermined values before measurement. The spin of a particle (or the polarization of a photon) follows these same patterns.

\paragraph*{Difference between zero gravity heads and tails and quantum heads and tails (Stern and Gerlach)}

The main difference relates to experiments 3 and 4 (Figures~\ref{fig:2appareils_b} et \ref{fig:3appareils}) where at least one second measurement is carried out in an orientation perpendicular to the first one. For these experiments, the result of the second measurement (Figure~\ref{fig:2appareils_b}) gives a 50\% chance of obtaining heads and a 50\% chance of obtaining tails, because the second measurement is perpendicular to the first. If it were not perpendicular, that is to say oriented by 90° in relation to the first hand clap, but merely oriented by 45°, for example, then the second measurement would in all cases yield the same result as the first. In fact, the hand clap cannot flip over the coin but simply straighten it in the axis of the hands. This results in a 50/50 chance of heads or tails because the orientation of the two measuring devices is 90° and therefore the coin is on its side during the second measurement. In quantum heads or tails, each orientation between the two measuring devices corresponds to a different up/down frequency of occurrence. All up/down frequencies of occurrence are possible. The frequency 50/50 is found when the orientation between the two measurement devices is perpendicular.

Although the question of the non-contextuality of the measurement of the game of heads or tails in space does not pose any problem of interpretation, the question of the non-contextuality of the spin measurement is the source of the divergences that exist among the main interpretations of quantum mechanics.
\begin{itemize}
 \item The Copenhagen interpretation (orthodox interpretation) refutes the existence of the spin orientation before its measurement by an observer; it is the measurement that creates the spin orientation; before measurement, the spin orientation is indeterminate. This assumption is extended to other properties such as a particle's position or velocity, which are assumed to be non-existent until they are measured.
 \item Everett's multiple worlds interpretation explains non-contextuality by splitting the world in two; each new world has a different context that gives a different result.
 \item In the de Broglie-Bohm interpretation, the non-contextuality of the spin measurement is explained in the same way as the non-contextuality of the measurement in the game of heads or tails in space: the measuring device guides the spin (resp. the coin orientation) to a discrete value, but the spin orientation (resp. the coin orientation) exists throughout the experiment.
\end{itemize}
Let us recall that these three approaches share the same equations (those of quantum mechanics) and therefore have the same physical predictions. All that distinguishes them is the interpretation of the equations.

In the following section, we present the dBB interpretation, which extends the analogy of the interpretation of the measurement of heads or tails in zero gravity to "the quantum world".

\section{ Spin "measurement" in Broglie-Bohm interpretation}
\label{sect:dBB}
Finally, we recall how the de Broglie-Bohm interpretation provides a simple explanation for the spin "measurement" according to the different axes. The measurement of the spin of a silver atom is carried out by a Stern-Gerlach  apparatus: an electromagnet $\mathbf{A_1}$, where there is a strongly inhomogeneous magnetic field $\textbf{B}$, followed by a screen $\mathbf{P_1}$ (Fig.\ref{fig:SG_schema}). The magnetic field $\textbf{B}$ is directed along the $Oz$ axis and the atomic beam is directed along the $Oy$ axis.

\subsection{Point source or spatial extension: spinor}

\begin{figure}
\begin{center}
\includegraphics[width=0.25\linewidth]{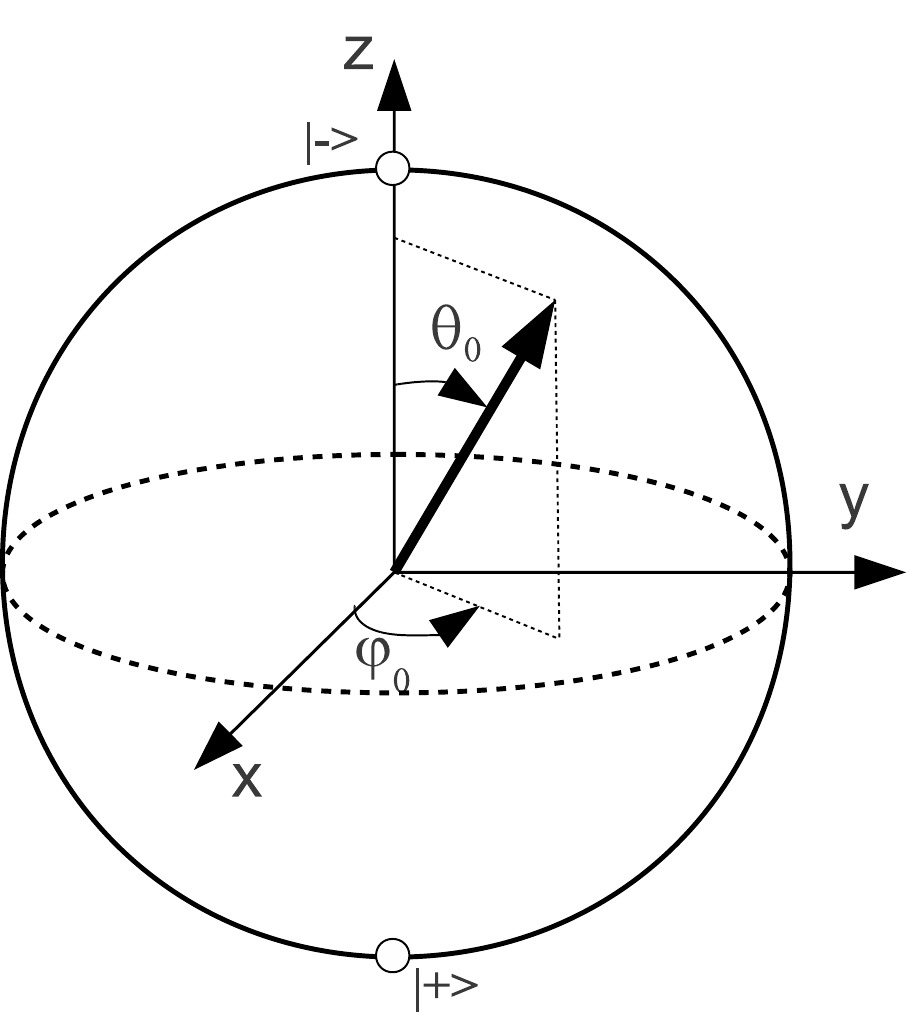}
\caption{\label{fig:spin} Orientation of the spin of a pure state. $\theta_0$ and $\varphi_{0}$ are the polar angles characterizing the spinor. They correspond to the angles of the Bloch sphere.}  
\end{center}
\end{figure}

The state of a spin 1/2 particle is given by the wave function $\Psi(\textbf{x},t)$, called a Pauli spinor, which has two complex components $\Psi^{+}(\textbf{x},t) $ and $\Psi^{-}(\textbf{x},t) $. At the moment of entering the magnetic field ($t=0$), we can associate to each silver atom a spinor corresponding to a pure state, as in many textbooks of quantum mechanics~\cite{Feynman1965, Cohen1977, Sakurai2011}~:
\begin{equation}\label{eq:psi0_ses}
    \Psi_{0} = \Psi(t=0) = \left( \begin{array}{c}\cos \frac{\theta_0}{2}e^{ i\frac{\varphi_0}{2}}
                                   \\
                                  \sin\frac{\theta_0}{2}e^{- i\frac{\varphi_0}{2}}
                  \end{array}
           \right)
\end{equation}
where $\theta_0$ and $\varphi_{0}$ are the polar angles characterizing the pure state (cf. figure~\ref{fig:spin}). The angles $\theta_0$ and $\varphi_{0}$ are unknown and each atom of the beam has angles $(\theta_0,\,\varphi_{0})$ which a priori are different; 
we only suppose that $\theta_0$ is uniformly distributed on $[0;\,\pi]$ and $\varphi_{0}$ on $[0;\,2\pi[$. The beam passing through the magnetic field is a statistical mixture of pure states, that is, a random mixture of atoms in different spinors.

To analyze the evolution of the spinor during the experiment and to understand how the spin is measured, it is absolutely necessary to represent an individual particle with spin by a spinor with a spatial extension as follows: 
\begin{equation}\label{eq:psi0_aes}
    \Psi_{0}(x,z) = (2\pi\sigma_{0}^{2})^{-\frac{1}{2}}
                      e^{-\frac{x^2+z^2}{4\sigma_0^2}}
                      \left( \begin{array}{c}\cos \frac{\theta_0}{2}e^{ i\frac{\varphi_0}{2}}
                                   \\
                                  \sin\frac{\theta_0}{2}e^{-i\frac{\varphi_0}{2}}
                  \end{array}
           \right)
\end{equation}
We consider the beam of silver atoms as a Gaussian beam along the $Oz$ and $Ox$ axes of standard deviation $\sigma_0=10^{-4}m$. The atomic propagation axis $Oy$ is treated classically, knowing that the velocity of the atoms in this direction is approximately $v_y = 500m/s$.

\subsection{Pauli equation}
 
 The evolution of the spinor  $\Psi(\textbf{x},t)=\binom{\Psi^{+}(\textbf{x},t)}
                            {\Psi^{-}(\textbf{x},t)}$, for a neutral spin-1/2 particle (silver atom) is given by the Pauli equation:
\begin{equation}\label{eq:Pauli}
    i\hbar \frac{\partial \Psi}{\partial t}(\textbf{x},t)
    =\left(-\frac{\hbar ^{2}}{2m}\boldsymbol\nabla^{2}
     +\mu_B \mathbf{B}\boldsymbol\sigma\right) \Psi(\textbf{x},t)
\end{equation}
where $m = 1.8\times 10^{-25} kg$ is the the mass of the silver atom, $\mathbf{B}$ is the magnetic field, $\mu_B=\frac{e\hbar}{2m_e}$ is the Bohr magneton and $\boldsymbol\sigma=(\sigma_{x},\sigma_{y},\sigma_{z})$ corresponds to the three Pauli matrices.

Without the initial spatial extension of the spinor (equation~\ref{eq:psi0_ses}), the spatial resolution of the Pauli equation is impossible and we lose the possibility of taking into account the evolution of the spin during the measurement process as shown by Takabayasi,~\cite{Takabayasi1955} Bohm et al.~\cite{Bohm1955a} and Dedwey et al.~\cite{Dewdney1986} 
Indeed, the differing evolutions of the two components of the spinor play a key role in understanding the measurement process.

\subsection{Solution without Pauli's equation and with measurement postulates}

Using measurement postulates, it is very easy to infer the result of the measurement without using the Pauli equation. We rewrite the initial spinor (equation~\ref{eq:psi0_ses}) with no spatial extension $\Psi_{0}$ (the reasoning is similar whether or not spatial extension is considered) in the form:
\begin{equation}\label{eq:psi02}
    \Psi_{0} = \cos\frac{\theta_0}{2}e^{i\frac{\varphi_0}{2}}|+\rangle
              +\sin\frac{\theta_0}{2}e^{-i\frac{\varphi_0}{2}}|-\rangle,
\end{equation}
where $|+\rangle=\binom{1}{0}$ and $|-\rangle=\binom{0}{1}$ are the eigenvectors of the operator $\sigma_z$. 

From the measurement postulates, $|\cos\frac{\theta_0}{2}e^{i\frac{\varphi_0}{2}}|^2=\cos^2\frac{\theta_0}{2}$ is the probability of measuring the spin $+\hbar/2$ (the eigenvalue of the projection operator $S_z =\frac{\hbar}{2}\sigma_z$) and $\sin^2\frac{\theta_0}{2}$ of measuring the spin $-\hbar/2$. The $\theta_0$ is unknown at the output of the source and the probability of measuring the spin $1/2$ when we do not know the initial spinor of the incident atom is equal to: $\int_0^\pi\cos^2\frac{\theta_0}{2}d\theta_0=0.5$. The probability of measuring the spin $-1/2$ is equal to: $\int_0^\pi\sin^2\frac{\theta_0}{2}d\theta_0=0.5$. We thus find the two distinct spots of identical intensity $N^+$ and $ N^-$. 

It is easy to understand the practical relevance of the postulates of the measurement: they allow the experimental results to be retrieved simply and quickly. However, they do not provide an accurate understanding of the measurement phenomenon.

\subsection{The de Broglie-Bohm theory}

In the de Broglie-Bohm (dBB) theory, the wave function does not completely represent the state of a quantum particle and it is necessary to add to its description the position of the particle $X(t)$. 
The evolution of the spinor is always given by the Pauli equation~(\ref{eq:Pauli}) and the evolution of the position is given by the following equation~\cite{Bohm1955a,Takabayasi1955}: 
\begin{equation}\label{eq:vitesse}
\dfrac{dX(t)}{dt}=\frac{\hbar}{2m \rho} Im{(\Psi^\dag\boldsymbol\nabla \Psi)}\vert_{\textbf{x}=X(t)}
\end{equation}	
where $\Psi^\dag=(\Psi^{+*}, \Psi^{-*})$ and $\rho=\Psi^\dag\Psi$. The atom is both a wave \emph{and} a particle. Wave and particle coexist, the first guiding the second by the equation~(\ref{eq:vitesse}), hence the name of "pilot wave" given by de Broglie in 1927~\cite{deBroglie1927}.
Bohm et al.\cite{Bohm1955a} also define $\mathbf{s}$ a spin vector field as: 
\begin{equation}\label{eq:spinvector}
\mathbf{s}(\mathbf{x},t)= \frac{\hbar}{2\rho}\Psi^\dag(\textbf{x},t)\sigma\Psi(\textbf{x},t)=\frac{\hbar}{2}(sin\theta~ sin\varphi, sin\theta ~cos\varphi, cos\theta).
\end{equation}	
The spin vector of an individual particle is evaluated along its trajectory as ~ $\textbf{s}= \textbf{s}(X(t),t)$. This spin vector is totally defined by the spinor and the position of the particle.

\subsection{Pure state}
 
We solve the Pauli equation for a single atom corresponding to a pure state with $(\theta_0,\ \varphi_0)$.
In the free space, at the end of the field, at the time $t+\Delta t$ ($t\geq0$ and $\Delta t=\Delta l/v_y$ being the crossing time of the field), one obtains~\cite{Gondran2005b}~:
\begin{equation}\label{eq:apreschamp}
\Psi(x,z,t+\Delta t)\simeq  (2\pi\sigma_0^2)^{-\frac{1}{2}}e^{-\frac{x^2}{2\sigma_0^2}}
\left(
\begin{array}{c}
                                \cos \frac{\theta_0}{2}
                 e^{-\frac{(z-z_{\Delta}- ut)^2}{2\sigma_z^2}}
                  e^{i\frac{m u z + \hbar \varphi_+}{\hbar }} \\
                                i \sin \frac{\theta_0}{2} 
                     e^{-\frac{(z+z_{\Delta}+
                  ut)^2}{2\sigma_z^2}}
e^{i\frac{-
    muz + \hbar \varphi_-}{\hbar }}
                            \end{array}
                     \right)
\end{equation}
where
\begin{equation}\label{eq:zdeltavitesse}
    z_{\Delta}=\frac{\mu_B B'_{0}(\Delta
    t)^{2}}{2 m}=10^{-5}m,~~~~~~u =\frac{\mu_B B'_{0}(\Delta t)}{m}=1 m/s,~~~~~~\mathbf{B}=\left(
\begin{array}{c}B'_0x\\0\\B_0-B'_0 z\end{array}\right),
\end{equation}
and where $\varphi_-$ et $\varphi_+$ are constants. 

\begin{figure}
\begin{center}
\includegraphics[width=0.5\textwidth]{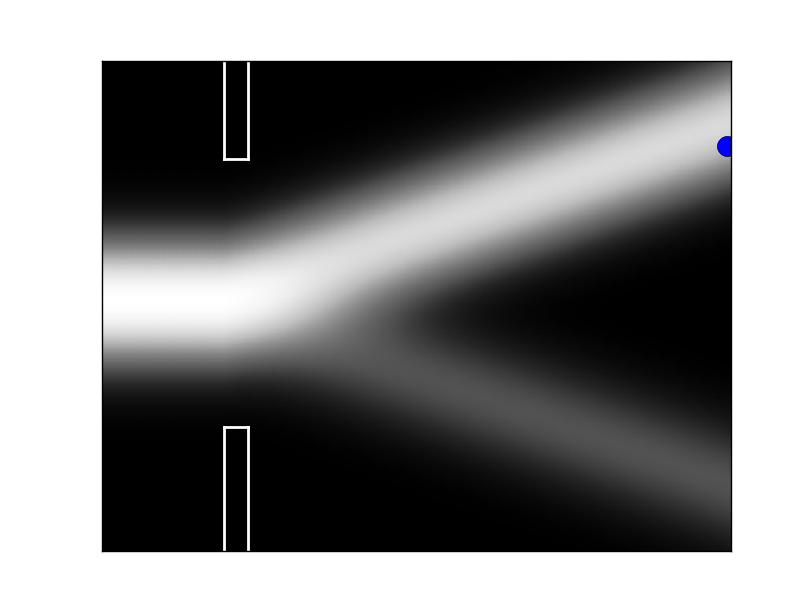}
\caption {\label{fig:etat_pur_a} Orthodox quantum mechanics: Pauli equation + reduction of the wave packet (final position). The notion of trajectory does not exist; only the position at the time of the measurement is defined using the postulate of reduction of the wave packet. The atom is either a wave (during all the experiment) or a particle (at the moment of impact).}
\end{center}
\end{figure}

\begin{figure}
\begin{center}
\includegraphics[width=0.5\textwidth]{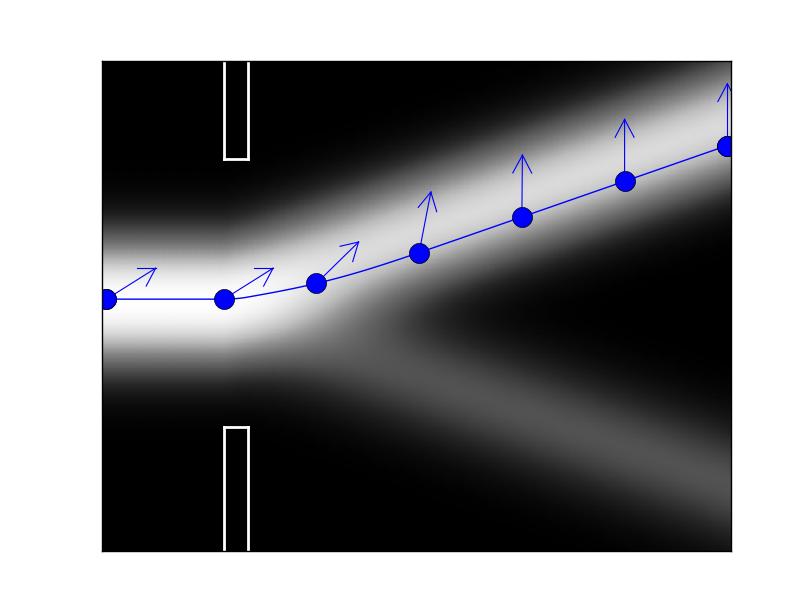}
 \caption{\label{fig:etat_pur_b} dBB theory (quantum mechanics taken one step further): Pauli equation + initial position.
The arrows indicate the $\theta$ orientation of the spin vector $\mathbf{s}$ (initially $\theta_0=\pi/3$).
The position of the particle exists before the measurement;
the particle follows a deterministic trajectory and the impact on the screen only reveals its position. The atom is both a wave \emph{and} a particle during all the experiment.}
\end{center}
\end{figure}

Figures~\ref{fig:etat_pur_a} and ~\ref{fig:etat_pur_b} represent the evolution of $\rho(z,t)=\int\Psi^\dag(x,z,t)\Psi(x,z,t)dx$, the probability density of the presence of the silver atom in a pure state for the values $\theta_0=\pi/3$ and $\phi_0=0$. The axis $Oy$, along which the beam is propagated, is on the abscissa ($y=v_y t$) and the axis $Oz$ is on the ordinate (the variable $x$ is not represented because the wave remains Gaussian according to this axis). The magnet $\mathbf{A_1}$ is represented in the figure by the two white rectangles, its length is $\Delta l = 1cm$ and there is $ D=20cm $ of free path before the atom is measured on the detection screen $\mathbf{P_1}$.
The numerical data comes from the textbook by Cohen-Tannoudji et al.~\cite{Cohen1977}. For the electromagnetic field $\textbf{B}$, we have: $B_{0}=5$ Tesla, $B'_{0}=\left| \frac{\partial B}{\partial z}\right| =- \left| \frac{\partial B}{\partial x}\right|= 10^3$ Tesla/m.

The dBB theory provides a more detailed explanation than the usual quantum mechanics explanation because, as seen in figure ~\ref{fig:etat_pur_b}, it shows how the \textit{initial} position of the particle explains in a causal way the \textit{final} position usually justified by the reduction of the wave packet (figure~\ref{fig:etat_pur_a}). In this framework, there is no measurement problem because the continuation of quantum mechanics towards classical mechanics is done in a natural way by the position of the particle.

\begin{figure}
\begin{center}
   \includegraphics[width=.5\textwidth]{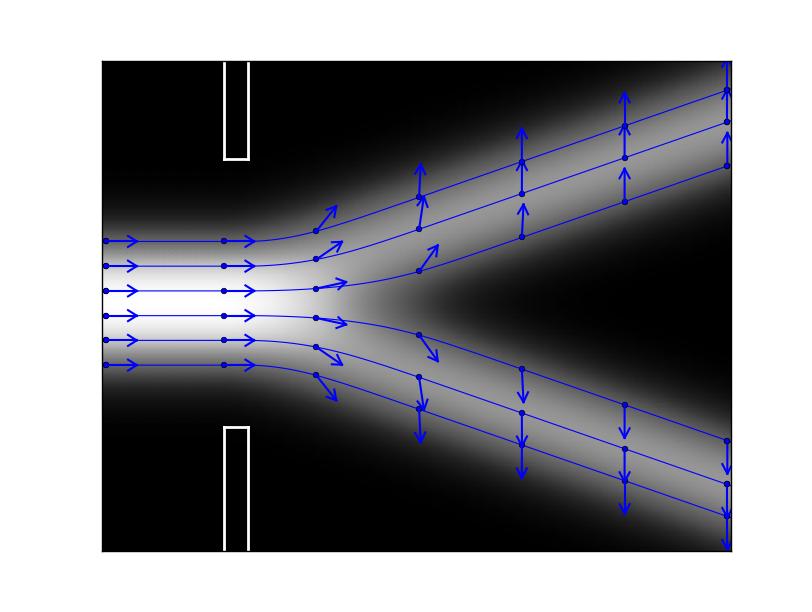}\ 
 \caption{\label{fig:etat_pur_traj} Six trajectories of silver atoms from the same pure state $\theta_0=\pi/2$.
The arrows indicate the evolution of $\theta$, initially all $\theta=\pi/2$.}
\end{center}
\end{figure}

Figure~\ref{fig:etat_pur_traj} represents the evolution of the density $\rho(z,t)$ for the pure state $\theta_0=\pi/2$ as well as six trajectories of atoms whose initial position $z_0$ was drawn at random.

The Pauli equation  only provides the probability $\rho$ of the impact of the atom.

Experimentally, one does not directly measure the spin of an atom, but $(\widetilde{x}$, $\widetilde{z})$ the position of the impact of the atom on $\mathbf{P_1}$.
 
If $\widetilde{z}\in N^+$, the term $\psi^-$ of~(\ref{eq:apreschamp}) is numerically zero and the spinor $\Psi$ is proportional to $\binom{1}{0}$,
one of the eigenvectors of $\sigma_z$. If $\widetilde{z}\in N^-$, the term $\psi^+$ of~(\ref{eq:apreschamp}) is numerically zero and the spinor $\Psi$ is proportional to $\binom{0}{1}$, the other eingenvector of $\sigma_z$. Therefore, the spin measure corresponds to an eigenvalue of the spin operator $S_z= \frac{\hbar}{2}\sigma_z$. 
It is a demonstration of the postulate of quantification by the dBB theory.

Equation~(\ref{eq:apreschamp}) gives the probabilities $\cos^{2}\frac{\theta_0}{2}$ (resp. $\sin^{2}\frac{\theta_0}{2}$) to measure the particle in the spin state $+\frac{\hbar}{2}$ (resp. $-\frac{\hbar}{2}$). It is a demonstration of the postulate of spatial decomposition by the dBB theory. The "measured" value is not a pre-existing value. It is contextual and conforms to the Kochen and Specker theorem \cite{Kochen1967}.

\begin{figure}
\begin{center}
  \includegraphics[width=.5\textwidth]{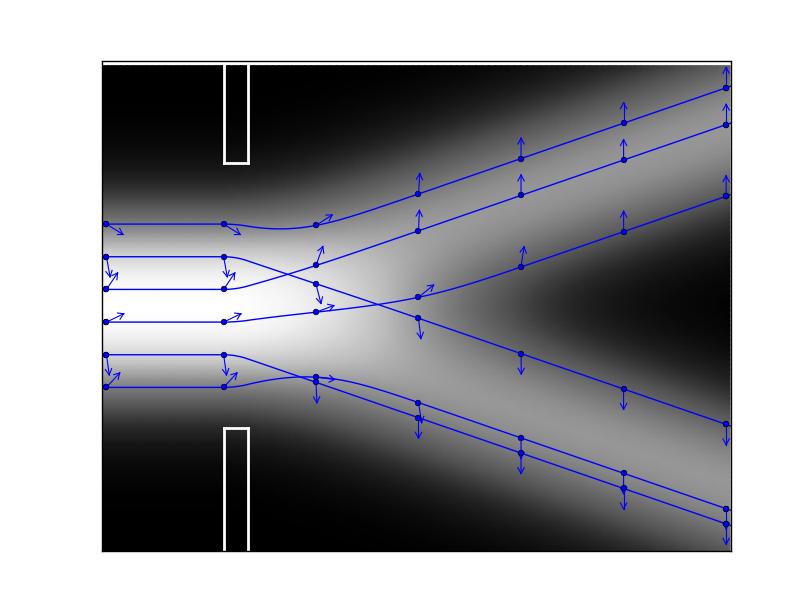}
 \caption{\label{fig:melange_traj}Six trajectories of silver atoms resulting from a statistical mixture of pure states.
The initial arrows indicate the $\theta_0$ of the pure state (i.e. of the spinor)
of each particle.}
\end{center}
\end{figure}

\subsection{Statistical mixture of pure states}

Since the beam of silver atoms is a mixture of pure states, the atomic density $\rho(z,t+\Delta t)$  is determined by integrating $\rho(x, z,t+\Delta t)$ on $x$ and on $(\theta_0,\varphi_0)$; we find:
\begin{eqnarray}\label{eq:apreschamp_melange}
    \rho(z,t+ \Delta t) &=
     (2\pi\sigma_0^2)^{-\frac{1}{2}}
                  \frac{1}{2}\left(e^{-\frac{(z-z_{\Delta}- ut)^2}{2\sigma_0^2}}+
                  e^{-\frac{(z+z_{\Delta}+
                  ut)^2}{2\sigma_0^2}}\right).
\end{eqnarray}
Figure~\ref{fig:melange_traj} represents this density $\rho(z,t)$ as well as  six trajectories of atoms whose initial position $z_0$  and the orientation of the initial spinor $\theta_0$ are drawn at random. We note the difference with the results of figure~\ref{fig:etat_pur_traj} for the pure state $\theta_0=\pi/2$ which gives statistically the same results. For a statistical mixture of pure states, the trajectories of the atoms of a beam can cross.

\section{Conclusion}
\label{sect:conclusion}
By introducing the game of heads or tails in weightlessness, we have shown that there is a contextual measure in classical mechanics and that there are no mysteries.
The "measured" value is not a pre-existing value. It is discretized by the measuring device (the clapping of hands). We find the same property for the "measurement" of the spin in de Broglie-Bohm's interpretation of the Stern and Gerlach experiment.

The "measurement" of the spin along the $Oz$ axis corresponds to the rectification of the orientation of the spin either in the direction of the gradient of the magnetic field or in the opposite direction. The result depends on the position of the particle in the wave function. The duration of the measurement is the time required for the particle to straighten its spin in the final direction. The "measured" value (the spin) is not a pre-existing value such as the mass and the charge of the particle but a contextual value conforming to the Kochen and Specker theorem.

We have also shown \cite{Gondran2016} that, for the two entangled particles of the EPR-B experiment, it is possible to replace the singlet spinor in configuration space by two single-particle spinors in physical space. The "measurement" of the two spins is contextual and verifies Bell's inequalities. The wave function of the singlet state alone introduces a non-local influence on the  the spin orientation, but not on the motion of the particles themselves. This is a key point in the search for
a physical understanding of this non-local influence.

\bibliographystyle{elsarticle-num}
\bibliography{biblio_mq.bib}

\end{document}